\documentclass[%
twocolumn,
showpacs,
nofootinbib,
nobibnotes,
amsmath,amssymb,
prd,
pra,
longbibliography,
floatfix,
]{revtex4}

\usepackage{amsmath}
\usepackage{mathrsfs}
\usepackage{amssymb}
\usepackage{graphicx}
\usepackage{etoolbox}
\usepackage{amsmath}
\usepackage{mathrsfs}
\usepackage{amssymb}
\usepackage{graphicx}
\usepackage{etoolbox}
\usepackage{dcolumn} 
\begin{document}\emph{}
	\preprint{APS/123-QED}
	
	\title
	{Masses and Mixing of Neutral Leptons in a Grand Unified $E_{6}$ Model with Intermediate Pati-Salam Symmetry}
	
	\author{Se\c{c}il Benli }
	\affiliation{Department of Physics, Ko\c{c} University, Sar{\i}yer, Istanbul, 34450, Turkey}
	
	\author{Tekin Dereli}
	\affiliation{Department of Physics, Ko\c{c} University, Sar{\i}yer, Istanbul, 34450, Turkey}
	\email{tdereli@ku.edu.tr}
	
	\date{\today}

\begin{abstract}

A brief review of the assignment of elementary fermions and bosons to irreducible multiplets  in grand unified $E_6$ models is 
followed by a discussion of 
different, hierarchical symmetry breaking chains from $E_6$ down to $SU(3)_C \times U(1)_{EM}$. 
We concentrate here on a model with an intermediate Pati-Salam symmetry for which $(B-L)$ is conserved. 
In particular, the mass/mixing matrix of electrically neutral fermions (i.e.neutrinos)  that would be derived from Yukawa couplings is constructed. 
The pattern of neutrino masses and some bounds  on mixing parameters are discussed. 

\end{abstract}

\maketitle

\section{Introduction} \label{sec:introduction}

\noindent The modern theory of high energy physics is based on the Standard Model which is a local quantum gauge field theory supporting the Lie symmetry algebras
$SU(3)$, $SU(2)$ and $U(1)$ that are associated with  strong, weak and electromagnetic interactions among the elementary particles.
It  is augmented with a fine-tuned Higgs mechanism that generates the  observed mass spectrum of all constituent fermions and the intermediate vector bosons. 
The existence  of an elusive heavy scalar Higgs boson in this framework was eventually confirmed at LHC experiments in 2012.

There are many reasons for not  
regarding the Standard Model as a final theory. First of all, even though it agrees very well with the experimental data available so far, it has too many independent 
free parameters to be fixed phenomenologically. This asks for an enlargement of the symmetry schemes in order to cut down the number of free parameters. 
Secondly, there is more in the Nature then that is evidenced by the Standard Model, because 
(i) current inflationary cosmological evolution scenarios require the existence of large amounts of Dark Matter and Dark Energy and (ii) the observed neutrino oscillations
imply massive neutral fermions with complicated patterns of interactions. Moreover, (iii) the fact that only three families of elementary particles exist 
still awaits for a theoretical explanation.  Finally, (iv) the puzzle of the matter-anti-matter asymmetry of the Universe also requires an explanation. 
These are the main reasons why we need new physics which holds out more symmetry beyond the Standard Model.
The new physics on which we will concentrate here is based on the idea of Grand Unification where three types of known gauge interactions are unified at the same very high energy scale by linking the three of the four fundamental  forces in nature  by combining strong and electroweak forces but excluding gravity.

Arguably the best known GUT (Grand Unification Theory) is due to Georgi and Glashow\cite{georgi1974unified}  that is based on the block-diagonal embedding 
$$
(U(1) \times SU(2))_{EW} \times  SU(3)_C  \hookrightarrow SU(5).
$$
$SU(5)$ is a rank-4 Lie group that supports a single (running) coupling constant. The grand unification may occur at extremely high energy scales at $\sim 10^{15} GeV$\cite{georgi1974hierarchy}. 
The broken symmetry pattern at lower energy scales is achieved through a two step hierarchy 
\begin{eqnarray}  
SU(5) &\rightarrow &(U(1) \times SU(2))_{EW} \times  SU(3)_C \nonumber \\
  &\rightarrow &U(1)_{QED}  \times  SU(3)_C. 
\end{eqnarray}
The rank drops by one at the second step. This gives rise to a $U(1)_{Y}$ charge associated with the hypercharge quantum number $Y$.
The ultimate price to pay for grand unification is the necessity of proton decay. Such lepto-quark processes would probe physics almost at the Planck scales. The present-day observational lower limit on the proton life-time independent on the decay mode stands at $\gtrsim 10^{34}$ years
and poses a great challenge for future experiments \cite{gell1978color},\cite{langacker1981grand},\cite{pati2000proton},\cite{senjanovic2003proton}.

There are two  other popular GUT's   that historically followed $SU(5)$  soon after;
one based on the rank-5 Lie group $SO(10)$\cite{georgi1975state},\cite{fritzsch1975unified}  and the other on rank-6 exceptional Lie group $E_6$\cite{gursey1976universal},\cite{shafi1978e6},\cite{achiman1978quark},\cite{barbieri1980exceptional}. The symmetry breaking hierarchy for $SO(10)$ GUT of Fritzsch and Minkowski\cite{fritzsch1975unified} reads
\begin{eqnarray}
SO(10) \rightarrow SU(5) & \rightarrow& (U(1) \times SU(2))_{EW} \times  SU(3)_C   \nonumber \\
& \rightarrow& U(1)_{QED}  \times  SU(3)_C.  
\end{eqnarray}
The drop by one in rank at the final step is again associated with $U(1)_{Y}$ while the  drop by one in rank at the first step is associated with a new $U(1)^t$ charge. 

\noindent The symmetry breaking pattern for the Model I $E_6$ GUT of G\"{u}rsey, Ramond and Sikivie\cite{gursey1976universal},\cite{gursey1977mixings}, 
\cite{gursey1980patterns},
\cite{gursey1981higgs351} on the other hand goes as 
\begin{eqnarray}
E_6 \rightarrow SO(10) &\rightarrow & SU(5)  \nonumber \\
& \rightarrow & (U(1) \times SU(2))_{EW} \times  SU(3)_C  \nonumber \\
&\rightarrow & U(1)_{QED}  \times  SU(3)_C. 
\end{eqnarray}
Here the first two steps involve two separate $U(1)$ charges. 
We note that  both of these GUTs are anomaly-free and left-right symmetric. 

\noindent  A slightly different approach is adopted in one of the earliest grand unified models  proposed by Pati and Salam\cite{pati1974lepton},\cite{mohapatra1975natural},\cite{senjanovic1975exact},\cite{pati1978neutral}. It  involves the following embedding of the colour symmetry
$$
SU(3)_C \hookrightarrow SU(4)_C
$$
and treats lepton number as a fourth colour. 
Accordingly  a left-right symmetric extension of the   SM  is determined through the embedding
\begin{eqnarray}
&SU(3)_{C}&\times SU(2)_L \times U(1)_Y  \nonumber \\
 &\hookrightarrow &SU(4)_C \times SU(2)_L \times SU(2)_R  \nonumber \\
&\cong &SO(6) \times SO(4). \nonumber
\end{eqnarray}
This alternative route will be referred to as  Pati-Salam symmetry in what follows. 
If  we now further embed $SO(10) \hookrightarrow E_6$,  we consider the hierarchy
\begin{eqnarray}
E_6 &\rightarrow & SO(10) \rightarrow SU(2)_{L} \times SU(2)_{R} \times SU(4)_{C}  \nonumber \\
 &\rightarrow& U(2)_{EW} \times SU(3)_{C}  \nonumber \\
 &\rightarrow & U(1)_{QED} \times SU(3)_{C}.
\end{eqnarray}
Here a basic fermion multiplet in each generation becomes $\{27\}$ of $E_6$ rather than $\{16\}$ of $SO(10)$, thus implying  eleven extra new types of fermions\cite{ruegg1979masses},\cite{georgi1979masses},\cite{barbieri1980hierarchical}. 
Such a Grand Unified $E_6$ model with intermediate Pati-Salam symmetry  thus combines the leptons and quarks in each generation into a single fundamental representation of $E_6$   and allows for interactions among them in  such a way that the baryon number on its own is not conserved but the difference between the baryon and lepton numbers $ (B-L)$ would be conserved\cite{weinberg1979baryon},\cite{mohapatra1980},\cite{paschos1995baryon},\cite{harada2003hypercharge},\cite{mohapatra2014old}.  The most recent work on grand unification based on non-commutative geometries provides further motivation that supports the Pati-Salam approach\cite{chamseddine2013beyond},\cite{chamseddine2015grand}.

\medskip

\noindent In this paper we consider a Grand Unified $E_6$ model with intermediate Pati-Salam symmetry. In particular, we look at electrically neutral leptons (neutrinos)
and study their masses and mixing within each family. In section:2, the assignment of fermion multiplets is discussed in general. 
We make use of the computer program LieArt in Mathematica 
to work out the  Lie algebraic weights and the roots of $E_6$, given the subgroup decomposition.  The relevant projection operators for our symmetry breaking hierarchy are 
explicitly constructed. 
We also comment briefly on the spectrum of coloured quarks and charged leptons. 
 In section:3, a mass matrix for neutral leptons is written down and its eigenvalues and eigenvectors are studied in second order perturbation theory.
Furthermore certain bounds involving masses and mixing parameters are derived.
Concluding remarks are given in section:4.  

\bigskip

\section{Fermion Assignments}\label{sec:fermionassignments}

\noindent  In the $E_6$ model, bosons are assigned to the adjoint representation $\{78\}$ and fermions are assigned to the 
fundamental representation $\{27\}$. 
Therefore, in accordance with  the irreducible decomposition
\begin{equation}
\{27\} \times \{27\} = \{\bar{27}\} + \{351\} + \{\bar{351}\},
\end{equation}
it is possible to assign the Higgs scalar bosons to the representation $\{\bar{27}\}$ of $E_6$\footnote{We make the most economical Higgs multiplet assignment. Higher Higgs multiplets  $\{351\}$ and  $\{\bar{351}\}$ may also be considered, for example, to implement supersymmetry\cite{babu-bajc-susic}.}.
Our aim here is to determine the generic mass matrix of the fermions  in $\{27\}$ by examining the  invariant Yukawa couplings  
 via the symmetry breaking chain 
\begin{eqnarray}
E_{6} &\longrightarrow & SO(10)\times U(1)^{t} \nonumber \\
&\longrightarrow & SU(2)_{L}\times SU(2)_{R}\times SU(4) \nonumber \\
&\longrightarrow & SU(2)_{L}\times SU(2)_{R}\times SU(3)_{C}\times U(1)_{B-L} \nonumber \\
&\longrightarrow & U(1)_{Y} \times SU(2)_{I}\times SU(3)_{C}. \nonumber
\end{eqnarray}
To highlight  the fermionic  content predicted by our model, a good starting point would be the Cartan matrix of the GUT group $E_6$ for which  we write down all the simple roots 
and can get the information we need for any representation we are interested in. 
 We make use of the  computer program  {\sl LieArt}\cite{feger2015lieart} for the tensor decompositions, branching rules and some basic formulations of the $E_{6}$. 
The Cartan matrix of $E_{6}$ is explicitly written as
\begin{equation}
E_{6}=
\begin{pmatrix}
2 & -1 & 0 & 0 & 0 & 0 \\
-1 & 2 & -1 & 0 & 0 & 0 \\
0 & -1 & 2 & -1 & 0 & -1 \\
0 & 0 & -1 & 2 & -1 & 0 \\
0 & 0 & 0 & -1 & 2 & 0 \\
0 & 0 & -1 & 0 & 0 & 2
\end{pmatrix}.
\end{equation}
Each weight vector can be given as a linear combination of simple roots $\alpha_{i}$ as
$$
\Lambda = \sum_{i}\bar{\Lambda_{i}}\frac{2}{(\alpha_{i},\alpha_{i})}\alpha_{i}
$$
where $[\bar{\Lambda_{1}},..,\bar{\Lambda_{l}}]$ gives the weight in the dual basis. Dynkin indices of a weight $\Lambda$ are defined as
$$
a_{i}=2\frac{(\Lambda,\alpha_{i})}{(\alpha_{i},\alpha_{i})}=\sum_{j}\bar{\Lambda_{j}}\frac{2}{(\alpha_{j},\alpha_{j})}A_{ji}
$$
in the Dynkin basis. Then every irreducible representation is uniquely identified by an ordered set of integers $(a_{1},...,a_{l})$, and each such set is a highest weight of one and only one irreducible representation\cite{slansky1981group},\cite{georgi1999liealgebras},\cite{ramond2010group}. We are interested in the fundamental representation of $E_{6}$ where all the fermions in the SM and more appears in. We simply start by the highest weight and substract the simple roots and obtain the weight diagram for the representation. Highest weight for the $\{27\}$ of $E_{6}$ is $(100000)$.
 We can see below the weight diagram generated by LieArt\cite{feger2015lieart}:
 \bigskip
 
\noindent WeightSystem[Irrep[[E6][1,0,0,0,0,0]]

\noindent 1 0 0 0 0 0, \, -1 1 0 0 0 0, \, 0 -1 1 0 0 0, 

\noindent 0 0 -1 1 0 1, \, 0 0 0 -1 1 1, \, 0 0 0 1 0 -1, \, 0 0 0 0 -1 1, 

\noindent  0 0 1 -1 1 -1, \, 0 0 1 0 -1 -1,\, 0 1 -1 0 1 0,\, 0 1 -1 1 -1 0, \, 

\noindent 1 -1 0 0 1 0,\,-1 0 0 0 1 0,\,0 1 0 -1 0 0,\, 1 -1 0 1 -1 0, 

\noindent  -1 0 0 1 -1 0,\, 1 -1 1 -1 0 0,\, -1 0 1 -1 0 0,\, 1 0 -1 0 0 1, 

\noindent  -1 1 -1 0 0 1,\, 1 0 0 0 0 -1,\, -1 1 0 0 0 -1,\, 0 -1 0 0 0 1, 

\noindent 0 -1 1 0 0 -1, \, 0 0 -1 1 0 0, \, 0 0 0 -1 1 0, \, 0 0 0 0 -1 0 .   


\bigskip 

\noindent Since we are interested in physical content, we should look for color and flavor embeddings in the model. The Dynkin indices can be converted into eigenvalues of a set of diagonal generators with
$$
Q(\Lambda)=\sum_{i} \bar{Q_{i}}a_{i} .
$$ 
The electric charge operator measured in this way for $E_{6}$ is
$$
Q^{EM}=\frac{1}{3}[212010] .
$$
We should now concentrate on the subgroup structure and the related symmetry breaking chain to gather a detailed information about the particle content. $E_{6}$ GUT has physical interest when $SU(3)_{C}$ is contained as a subgroup.
In this work we concentrate on the chain where $SU(4)$ contains $SU(3)_{C}$. We follow the projection of highest weight of the representation to the highest weight it branches to. Thus we work with different projection matrices at each step of the symmetry breaking chain. For the chain $E_{6}\supset  SO(10)\supset SU(2)_{L}\times SU(2)_{R}\times SU(4)$ where $SU(4)\supset SU(3)_{C}$, we have the following projection matrices:
\[
P_{1}(E_{6}\supset SO(10))=
\begin{pmatrix}
0 & 1 & 1 & 1 & 0 & 0\\
0 & 0 & 0 & 0 & 0 & 1 \\
0 & 0 & 1 & 0 & 0 & 0 \\
0 & 0 & 0 & 1 & 1 & 0\\
1 & 1 & 0 & 0 & 0 & 0 \\
\end{pmatrix},
\]
where $P_{1}$ projects the weights to the subgroup $SO(10)$, and 
\begin{eqnarray}
P_{13}&(SO(10)&\supset SU(2)_{L}\times SU(2)_{R}\times SU(4)) \nonumber \\
&=&
\begin{pmatrix}
0 & 0 & 1 & 1 & 1 \\
0 & 0 & 1 & 0 & 0\\
1 & 1 & 1 & 0 & 1 \\
0 & 1 & 1 & 1 & 0\\
-1 & -1 & -1 & -1 & 0 
\end{pmatrix},
\end{eqnarray}

where $P_{13}$ projects the weights in $SO(10)$ to the weights in the $SU(2)_{L}\times SU(2)_{R}\times SU(4)$ representation, and 
\[
P_{5}(SU(4)\supset SU(3)_{C})=
\begin{pmatrix}
1 & 0 & 0 \\
0 & 1 & 0
\end{pmatrix},
\]
where $P_{5}$ projects the weights in $SU(4)$ to the weights in $SU(3)_{C}$. At every step of symmetry breaking chain we get different $U(1)$ charges.
We project $SO(10)$ weight to the subgroup $SU(2)_{L}\times SU(2)_{R}\times SU(4)$ and the first Dynkin index in the projected weight $(a_{1}a_{2}a_{3}a_{4}a_{5})$ gives the weak isospin $I_{3}^{L}=a_{1}/2$, the second Dynkin index gives  $I_{3}^{R}=a_{2}/2$ and the remaining indices  $(a_{3}a_{4}a_{5})$ fix the weight in $SU(4)$. Then we take this $SU(4)$ weight and project it once more relative to the subgroup $SU(3)_{C}$ and get the color weight in $SU(3)_{C}$. 
Therefore, the   modified Gell-Mann-Nishijima formula that gives the electric charge reads
\begin{equation}
Q^{EM} = I_3^L + I_3^R + \frac{1}{2}(B-L).
\end{equation}

\noindent  (A) The labelling and the physical interpretation of the corresponding (first generation) neutral fermions are as follows:
\begin{itemize}
\item $\nu_e$ is the left-handed  electron neutrino with quantum numbers $I_3 = 1/2, Y=-1, B-L=-1,$
\item $N_e^C $ is the charge-conjugate of right-handed  electron neutrino with quantum numbers $I_3 = 0, Y=0, B-L=1,$
\item $\nu_E$ is the left-handed  exotic neutrino with quantum numbers $I_3 = 1/2, Y=-1, B-L=0,$
\item $N_E^C$ is the charge conjugate of right-handed  exotic neutrino with quantum numbers $I_3 = -1/2, Y=1, B-L=0, $
\item $\nu_S$ is the left-handed  sterile neutrino with quantum numbers $I_3 = 0, Y=0, B-L=0. $
\end{itemize}

\begin{table}
	\center
	\begin{tabular}{ |c|c|c|c|c|c|c| }
		\hline
		\multicolumn{7}{|c|}{Quantum Numbers of Neutral Leptons} \\
		\hline
		Particle Assignment & $Q^{t}$ & $I_{3}^{L}$&$I_{3}^{R}$& $(B-L)$ & $Y$&$Q^{EM}$ \\
		\hline
		$\nu_{e}$   & 1&1/2&0    &-1&   -1&0\\
		$N_{e}^{C}$ &   1&0&-1/2  & 1   & 0&0\\
		$N_{E}^{C}$ &-2 & -1/2&1/2& 0&  1&0\\
		$\nu_{E}$   & -2& 1/2&-1/2&0 & -1&0\\
		$\nu_S$ &   4&0&0  & 0&0&0\\
		\hline
	\end{tabular}
	\caption{$U(1)$-charges for first generation neutral leptons in left-right symmetric symmetry breaking chain in $E_{6}$.}
	\label{Table :1}
\end{table}

\noindent Assuming that $\{\bar{27}\}$ dominates  the Higgs boson sector, we now look for  particles with charges in $\{\bar{27}\}$ representations that come with minus signs.
We write the charges in  $\{\bar{27}\}$ in  $(\nu_{e},N_{e}^{C},\nu_{E},N_{E}^{C},\nu_S)$ basis as \cite{robinett1982mass},\cite{rosner2014three},\cite{joglekar-rosner}: \\

\begin{tabular}{ c|ccccc }
	& $\underset{(1,-1,-1)}{\nu_{e}}$ & $\underset{(1,1,0)}{N_{e}^{c}}$ & $\underset{(-2,0,-1)}{\nu_{E}}$ & $\underset{(-2,0,1)}{N_{E}^{c}}$ & $\nu_S$ \\
	\hline
	$\underset{(1,-1,-1)}{\nu_{e}}$ & &$\underset{(2,0,-1)}{}$ & $\underset{(-1,-1,-2)}{}$ &$\underset{(-1,-1,0)}{}$ & \\
	$\underset{(1,1,0)}{N_{e}^{c}}$ & $\underset{(2,0,-1)}{}$ & & & $\underset{(-1,1,1)}{}$ & \\
	$\underset{(-2,0,-1)}{\nu_{E}}$ & $\underset{(-1,-1,-2)}{}$ & & & $\underset{(-4,0,0)}{}$ & $\underset{(2,0,-1)}{}$\\
	$\underset{(-2,0,1)}{N_{E}^{c}}$ & $\underset{(-1,-1,0)}{}$ &$\underset{(-1,1,1)}{}$  & $\underset{(-4,0,0)}{}$ & & $\underset{(2,0,1)}{}$\\
	$\nu_S$ & & & $\underset{(2,0,-1)}{}$ & $\underset{(2,0,1)}{}$ & \\
\end{tabular}

\bigskip

\noindent Empty entries above mean that the corresponding charges do not occur in the $\{\bar{27}\}$ representation\footnote{The alternative symmetry breaking chain
with intermediate $SU(5)$ symmetry instead of the Pati-Salam symmetry leads to a mass matrix with the same non-vanishing entries as above except  the entry $M_{13} \sim (-1,-1,-2)$ that vanishes. 
This is the case considered  in Rosner\cite{rosner2014three}.}.

\bigskip

\noindent  (B)  The labelling and the physical interpretation of the corresponding (first generation) colour-triplet quarks are as follows:
\begin{itemize}
\item ${\bf u}_L$ is the left-handed  part of  up-quarks, with quantum numbers $I_3 = 1/2, Y=1/3, B-L=1/3,$
\item ${\bf d}_L $ is the left-handed  part of down-quarks, with quantum numbers $I_3 = -1/2, Y=1/3, B-L=1/3,$
\item ${\bf D}_R^C$ is the charge conjugate of right-handed part of exotic quarks, with quantum numbers $I_3 = 0, Y=2/3, B-L=2/3,$
\item ${\bf u}_R^C$ is the charge conjugate of right-handed part of up-quarks, with quantum numbers $I_3 = 0, Y=-1/3, B-L=-1/3,$
\item ${\bf d}_R^C$ is the charge conjugate of right-handed part of down-quarks, with quantum numbers $I_3 = 0, Y=2/3, B-L=1/3, $
\item ${\bf D}_L$  is the left-handed part of exotic quarks,  with quantum numbers $I_3 = 0, Y=-2/3, B-L=2/3.$
\end{itemize}

\begin{table}[h]
	\center
	\begin{tabular}{ |c|c|c|c|c|c|c|c|  }
		\hline
		\multicolumn{7}{|c|}{Quantum Numbers  of Color Triplet Quarks} \\
		\hline
		Particle Assignment & $Q^{t}$ &$I_{3}^{L}$ & $I_{3}^{R}$& $(B-L)$ & $Y$ & $Q^{EM}$\\
		\hline
		${\bf u}_{L}$   & $1$ & $1/2$ & $0$ & $1/3$ &$1/3$ & $2/3$\\
		${\bf d}_{L}$ & $1$ & $-1/2$ & $0$ & $1/3$ & $1/3$ & $-1/3$\\
		${\bf D}^{C}_{R}$ & $-2$ & $0$ & $0$ & $2/3$ & $2/3$ & $1/3$\\
		${\bf d}^{C}_{R}$   &$1$ & $0$ & $1/2$ & $-1/3$ & $2/3$ & $1/3$ \\
		${\bf u}^{C}_{R}$ &  $1$ & $0$ & $-1/2$ & $-1/3$ & $-4/3$ & $-2/3$\\
		${\bf D}_{L}$ & $-2$ & $0$ & $0$ & $-2/3$ & $-2/3$ & $-1/3$ \\
		\hline
	\end{tabular}
	\caption{$U(1)$-charges for first generation, color triplet quarks in left-right symmetric symmetry breaking chain in $E_{6}$.}
	\label{Table :2}
\end{table}

\noindent  We also write down the allowed charges in $\{\bar{27}\}$, relative to the above basis of quarks as 


\begin{table}[h]
\scalebox{0.75}{
\begin{tabular}{ c|cccccc }
	&$\underset{(1,1,1/3)}{{\mathbf u}_L}$ & $\underset{(1,-1/3,-4/3)}{{\mathbf d}_L}$ & $\underset{(1,-1/3,1/3)}{{\mathbf D}^{C}_{R}}$ & $\underset{(-1,1.1/3)}{{\mathbf d}_{L}^{C}}$ &$\underset{(-2,2/3,2/3)} {{\mathbf u}_{L}^{C}}$ & $\underset{(-2,-2/3,-2/3)}{{\mathbf D}_{L}}$ \\
	\hline
	$\underset{(1,1,1/3)}{{\mathbf u}_L}$ & & & &  &$\underset{(2,0,-1)}{}$ & \\
	  $\underset{(1,-1/3,-4/3)}{{\mathbf d}_L}$ & & & $\underset{(-1,1,1)}{}$ & $\underset{(2,0,,1)}{}$ & &  \\
	 $\underset{(1,-1/3,1/3)}{{\mathbf D}^{C}_{R}}$ & &$\underset{(-1,1,1)}{}$ &  &  & $\underset{(-1,-1,-2)}{}$  &$\underset{(-4,0,0)}{}$ \\
 $\underset{(-1,1.1/3)}{{\mathbf d}_{R}^{C}}$ & & $\underset{(2,0,1)}{}$ &  & & & $\underset{(-1,-1,0)}{}$ \\
$\underset{(-2,2/3,2/3)} {{\mathbf u}_{R}^{C}}$ & $\underset{(2,0,-1)}{}$ & & $\underset{(-1,-1,-2)}{}$ & &  &  \\
 $\underset{(-2,-2/3,-2/3)}{{\mathbf D}_{L}}$&  & & $\underset{(-4,0,0)}{}$ & $\underset{(-1,-1,0)}{}$ & &  \\
\end{tabular}
}
\end{table}

\bigskip

\noindent (C)  The labelling and the physical interpretation of the corresponding (first generation) charged leptons are as follows:
\begin{itemize}
\item $e_L$ is the left-handed part of  electron, with quantum numbers $I_3 = 1/2, Y=-1, B-L=-1,$
\item $e_R^C $ is the charge conjugate of right-handed part of  electron, with quantum numbers $I_3 = 0, Y=2, B-L=1,$
\item $E_L$ is the left-handed part of exotic lepton with quantum numbers $I_3 = 1/2, Y=-, B-L=0,$
\item $E_R^C$ is the charge conjugate of right-handed part of exotic lepton with quantum numbers $I_3 = -1/2, Y=-1, B-L=0.$
\end{itemize}

\bigskip

\begin{table}[h]
	\center
	\begin{tabular}{ |c|c|c|c|c|c|c|c|  }
		\hline
		\multicolumn{7}{|c|}{Quantum Numbers of  Charged Leptons} \\
		\hline
		Particle Assignment & $Q^{t}$ &$I_{3}^{L}$ & $I_{3}^{R}$& $(B-L)$ & $Y$ & $Q^{EM}$\\
		\hline
		$e_L$   & $1$ & $-1/2$ & $0$ & $-1$ &$-1$ & $-1$\\
		$e_R^C$ & $1$ & $0$ & $1/2$ & $1$ & $2$ & $1$\\
		$E_L$ & $-2$ & $-1/2$ & $-1/2$ & $0$ & $-1$ & $-1$\\
		$E_R$   &$-2$ & $1/2$ & $1/2$ & $0$ & $1$ & $1$ \\
		\hline
	\end{tabular}
	\caption{$U(1)$-charges for electrically charged leptons in left-right symmetric symmetry breaking chain in $E_{6}$.}
	\label{Table :3}
\end{table}

\noindent  The allowed charges in $\{\bar{27}\}$, relative to the basis of charged leptons given above will be 

\begin{table}[h]
\scalebox{0.8}{
	\begin{tabular}{c|cccc}
		&$e(1,-1,-1)$& $e^{c}(1,1,2)$ & $E(-2,0,-1)$ & $E^{c}(-2,0,1)$ \\
		\hline
		$e(1,-1,-1)$ & -- & $(2,0,1)$ & $(-1,-1,-2)$ & $(-1,-1,0)$\\
		$e^{c}(1,1,2)$ & $(2,0,1)$ & -- & $(-1,1,1)$ & --\\
		$E(-2,0,-1)$ & $(-1,-1,-2)$ & $(-1,1,1)$ & -- & $(-4,0,0)$\\
		$E^{c}(-2,0,1)$& $(-1,-1,0)$ & -- & $(-4,0,0)$ & --\\
	\end{tabular}}
\end{table}


\bigskip

\section{Masses and Mixing of Neutral Leptons}\label{sec:massmatrix}

\noindent The spectrum of electrically neutral fermion fields in our $E_6$-model consists of three generations of  (i) active left-chiral and right-chiral neutrinos\cite{drewes2013right}, (ii) 
heavy exotic  (Dirac) neutrinos\cite{tosa1985exotic},\cite{rizzo1986exotic},\cite{rizzo1986phenomenology},\cite{feger2015grand}, and (iii)  light sterile  (Majorana) neutrinos\cite{rosner2014three},\cite{giunti2014sterile}. We won't be dealing with mixings among different generations in the present paper. 
We introduce a minimal seesaw mechanism below to separate out  the  observable left-chiral active  neutrinos  from the  right-chiral active neutrinos
which remain unobservable at current energy scales. 
Such a separation in mass is not predicted by the $E_6$ models, however, we follow a common practice\cite{ma1996neutrino},\cite{rosner2014three},\cite{senjanovic2013seesaw},\cite{ma2017seesaw} and implement a minimal seesaw mechanism by hand in order to remain consistent  with current neutrino observations. 
We associate  large Dirac masses with  exotic neutrinos. The sterile neutrino masses could be small or even may be set to zero at first approximation to remain 
in favour with recent inflationary big bang cosmology scenarios\cite{robinett1982mass},\cite{ma1996neutrino},\cite{king2013neutrino}.

\medskip

\noindent In order to set the notation in general,   let us consider a left-handed, 2-component complex spinor field $\psi_{L}$  and a  right-handed, 2-component complex spinor field $\psi_{R}$ that transform as $(1/2,0)$ and $(0,1/2)$ irreducible representations of the Lorentz group, respectively \cite{dreiner et al},\cite{cheng-kong}. Their charge conjugate fields 
$\psi_{L}^{C} =i\sigma_2 \psi_{L}^{*}$ and  $\psi_{R}^{C} =i\sigma_2 \psi_{R}^{*}$ carry opposite chiralities. A 4-component Dirac spinor field can be given by
$$
\Psi = \left ( \begin{array}{c}  \psi_{L} \\ \psi_{R} \end{array} \right ),
$$
thus transforming under $(1/2,0) \oplus(0,1/2)$. A Majorana spinor field is self-charge conjugate and  satisfies  $\psi_{R}^{C} = \psi_{L}$ (so that $\psi_{L}^{C} = \psi_{R}$ as well ).

\noindent We consider the following  formal expression for the Lagrangian density of a Dirac fermion:
\begin{eqnarray}
	{\cal{L}}_{Fermion} &=& Her(i \bar{\Psi} (\gamma \cdot \nabla) \Psi)+  i\bar{\Psi} {\cal{M}} \Psi  \nonumber \\
	&=&  Her(i \bar{\psi}_{L} (\sigma \cdot \nabla) \psi_{L}) + Her(i \bar{\Psi}_{R}^{C} (\sigma \cdot \nabla) \psi_{R}^{C})  \nonumber \\ 
	+ & iM_{L}& \bar{\psi}_{L} \psi_{L}  + iM_{R} \bar{\psi}_{R}^{C}  \psi_{R}^{C}  
	+ i M \left ( \bar{\psi}_{L} \psi_{R}^{C} +  \bar{\psi}_{R}^{C} \psi_{L} \right )  \nonumber
\end{eqnarray}
where $M_{L},M_{R}$ are two Majorana masses and $M$ is a Dirac mass. By convention, all the 2-spinors that appear in the Lagrangian density are taken as left-handed. 
If $\psi_{L}$ describes a left-handed neutrino field, then $\psi_{R}^{C}$ describes an (independent) left-handed anti-neutrino field. 
Then  we write down by inspection the following $2 \times 2$ mass matrix
$$
{\cal{M}} = \left ( \begin{array}{cc} M_{L}&M\\M&M_{R} \end{array}  \right ).
$$

\medskip 

\noindent  Keeping with the same conventions as above,  we describe all the masses and mixing of neutral fermions in our $E_6$-model   
by the following $5 \times 5$ real symmetric matrix: 
\begin{equation}
{\mathcal{M}} = \left ( \begin{array}{ccccc}  0& m & M_{13} & M_{14}& 0\\  m &M& 0& m_{24}& 0\\ M_{13} &0&0& M^{\prime}& m_{35}\\ M_{14} & m_{24}& M^{\prime} &0& m_{45}\\
0& 0&m_{35} &m_{45}&-m^{\prime \prime}\\         \end{array}    \right ) .
\end{equation}
We have introduced three mass parameters $m,M^{\prime},m^{\prime \prime}$ and  one minimal 
seesaw parameter $M$. 
Furthermore there are five mixing parameters
$M_{13}, M_{14}, m_{24}, m_{35}, m_{45}$. Here and in what follows, lower case $m$'s correspond to "small"  parameter values while the upper case $M$'s correspond to "large" ones\cite{robinett1982mass}. All the mass parameters are assumed positive. 
Let us remark here that if the symmetry breaking chain involves an intermediate $SU(5)$ symmetry, then the mass/mixing matrix cannot support the entry $M_{13}$ . Then
it would be set  to zero above which is the  case studied by Rosner\cite{rosner2014three}. In our problem though, the symmetry breaking chain passes through an intermediate 
stage with Pati-Salam symmetry so that $M_{13} \neq 0$. 

\noindent A diagonalisation of the above matrix to determine the exact mass eigenvalues  doesn't seem feasible. 
Nevertheless we demand that  the signs and magnitudes of the free parameters should be so chosen that there will be two real positive and three real negative eigenvalues,  
with one positive eigenvalue and one negative eigenvalue being  equal to each other in absolute value.
In such a case, all the masses assigned to the neutral fermions would be  real and positive.  On the other hand, since we do not have sufficient clue concerning the magnitude of the exotic and/or sterile neutrino masses and their mixings, no numerical estimates would be possible.  Then , we do the best we can and  evaluate below some perturbative 
expressions  for the mass eigenvalues.
The starting point for our model building process will be the lowest order neutrino mass matrix
$$
{\mathcal{M}}_{0} = \left ( \begin{array}{ccccc}  0& m &0 &0& 0\\  m &M& 0&0& 0\\ 0&0&0& M^{\prime}&0\\0& 0& M^{\prime} &0& 0\\
0& 0&0&0& -m^{\prime \prime}\\         \end{array}    \right ) .
$$
The upper left $2 \times 2$ block refers to active neutrinos where $m$ is a Dirac mass and $M$ is introduced to induce a minimal seesaw mechanism.  
The middle $2 \times 2$ block refers to an exotic fermion whose Dirac mass is $M^{\prime}$. The lower right corner refers to a single,  self-conjugate sterile neutrino
for which we introduce a Majorana mass $m^{\prime \prime}$.  
The eigenvalues of ${\mathcal{M}}_{0}$ are found to be
\begin{widetext}
$$
M_{+}=\frac{M}{2} + \sqrt{ (\frac{M}{2})^2 + m^2}, \quad -m_{-}=\frac{M}{2} - \sqrt{ (\frac{M}{2})^2 + m^2}, \quad M^{\prime}, \quad -M^{\prime}, \quad -m^{\prime \prime},
$$

with the corresponding eigenvectors
$$
|1> = \frac{1}{\sqrt{M_{+}^2 +m^2}} \left (\begin{array}{c} m\\M_{+}\\0\\0\\0  \end{array} \right ), \; |2> = \frac{1}{\sqrt{m_{-}^2 +m^2}} \left (\begin{array}{c} m\\-m_{-}\\0\\0\\0  \end{array} \right ), 
$$
$$ 
|3>  = \frac{1}{\sqrt{2}} \left (\begin{array}{c} 0\\0\\1\\1\\ 0  \end{array} \right ), \; |4> = \frac{1}{\sqrt{2}} \left (\begin{array}{c} 0\\0\\1\\-1\\ 0  \end{array} \right ), \; 
|5> = \left (\begin{array}{c} 0\\0\\0\\0\\1  \end{array} \right ).
$$
\end{widetext}
In first approximation, for $M >> m$, we have
$$
M_{+} \cong M , \quad m_{-} \cong \frac{m^2}{M} 
$$
with the corresponding eigenvectors
$$
\frac{1}{\sqrt{M^2 +m^2}} \left (\begin{array}{c} m\\M\\0\\0\\0  \end{array} \right ), \; \frac{1}{\sqrt{{({\frac{m^2}{M}})^2 +m^2}}} \left (\begin{array}{c} m\\ -\frac{m^2}{M} \\0\\0\\0  \end{array} \right ). 
$$
Therefore, the mass of the active right-handed neutrino $\sim M$, could be  "large" so that it is a heavy neutrino; while the mass of the active left-handed neutrino $\sim \frac{m^2}{M}$, remains "small" so that it is a light neutrino. 

\bigskip

\noindent We regard the mixings as  perturbation on the zeroth order mass matrix given above:
$$
{\mathcal{M}}_1 = {\mathcal{M}} - {\mathcal{M}}_0 .
$$
Then we calculate the following second order perturbative corrections to the mass eigenvalues:
\begin{eqnarray}
\Delta M_{+} &=& -\frac{|<1|{\mathcal{M}}_{1}|3>|^{2} }{M_{+} - M^{\prime}} -\frac{|<1|{\mathcal{M}}_{1}|4>|^{2} }{M_{+} + M^{\prime}} , \nonumber \\
\Delta m_{-} &=& -\frac{|<2|{\mathcal{M}}_{1}|3>|^{2} }{-m_{-} - M^{\prime}} -\frac{|<2|{\mathcal{M}}_{1}|4>|^{2} }{-m_{-} + M^{\prime}},  \nonumber \\
\Delta M_{+}^{\prime} &=& -\frac{|<3|{\mathcal{M}}_{1}|1>|^{2} }{M^{\prime}-M_{+}} -\frac{|<3|{\mathcal{M}}_{1}|2>|^{2} } {M^{\prime}+m_{-}},  \nonumber \\
\Delta M_{-}^{\prime} &=& -\frac{|<4|{\mathcal{M}}_{1}|1>|^{2} }{ - M^{\prime}-M_{+}} -\frac{|<4|{\mathcal{M}}_{1}|2>|^{2} }{- M^{\prime}+m_{-}},  \nonumber \\
\Delta  m^{\prime \prime}&=& -\frac{|<5|{\mathcal{M}}_{1}|3>|^{2} }{ - M^{\prime}} -\frac{|<5|{\mathcal{M}}_{1}|4>|^{2} }{M^{\prime}}, \nonumber 
\end{eqnarray}
where the relevant matrix elements turn out to be 
\begin{widetext}
{\small
$$
<1|{\mathcal{M}}_{1}|3> = \frac{m(M_{13} + M_{14}) + M_{+} m_{24}}{\sqrt{ 2(m^2 + M_{+}^2 )}}   , \quad   <1|{\mathcal{M}}_{1}|4> =  \frac{m(M_{13} - M_{14}) -M_{+} m_{24}}{\sqrt{ 2(m^2 + M_{+}^2) }}   ; 
$$
$$
<2|{\mathcal{M}}_{1}|3> =  \frac{m(M_{13} + M_{14}) - m_{-} m_{24}}{\sqrt{ 2(m^2 + m_{-}^2 )}}     , \quad   <2|{\mathcal{M}}_{1}|4> =  \frac{m(M_{13} - M_{14})  + m_{-} m_{24}}{\sqrt{ 2(m^2 + m_{-}^2 )}}    ; 
$$
$$
<3|{\mathcal{M}}_{1}|5> = \frac{m_{35}+m_{45}}{\sqrt{2}}     , \quad   <4|{\mathcal{M}}_{1}|5> =  \frac{m_{35}-m_{45}}{\sqrt{2}}   . 
$$ 
}
\end{widetext}

\noindent Then, for instance, the sterile neutrino mass, assuming $m^{\prime \prime}=0$ to zeroth order in perturbation theory, is given by
\begin{equation}
m_S \cong \frac{2 |m_{35}m_{45}|}{M^{\prime}},
\end{equation}
so that it is inversely proportional to the exotic neutrino mass. Its magnitude is controlled by the mixing of exotic and sterile neutrinos.

\bigskip

\noindent If we consider the mass matrix appearing with the $U(1)$ charges derived from our symmetry breaking chain which does not include the term M on the diagonal
and since it has no exact solution, we can follow different matrix methods to determine the mass eigenvalues approximately\cite{MatrixAnalysis}. One of our main assumptions is that the exotic mass $M^{\prime} \equiv M_{34}$ is much 
larger than the others. Then we can diagonalise the mass matrix with respect to it by a Jacobi transformation 
\[
J_{34}=
\begin{pmatrix}
1 & 0 & 0 & 0 & 0 \\
0 & 1 & 0 & 0 & 0 \\
0 & 0 & c & -s & 0 \\
0 & 0 & s & c & 0 \\
0 & 0 & 0 & 0 & 1\\
\end{pmatrix},
\]
and obtain the matrix ${\mathcal{M}}^{\prime} = J_{34}^{T}{\mathcal{M}} J_{34}$ as 
\begin{widetext}
\[
{\mathcal{M}}^{\prime}=
\left(\begin{smallmatrix}
0 & m_{12} & (M_{13}+M_{14})/\sqrt{2} & (M_{14}-M_{13})/\sqrt{2} & 0 \\
m_{12} & 0 & m_{24}/\sqrt{2} & m_{24}/\sqrt{2} & 0 \\
(M_{13}+M_{14})/\sqrt{2} & m_{24}/\sqrt{2} & M_{34} & 0 & (m_{35}+m_{45})/\sqrt{2} \\
(M_{14}-M_{13})/\sqrt{2} & m_{24}/\sqrt{2} & 0 & -M_{34} & (m_{45}-m_{35})7\sqrt{2}\\
0 & 0 & (m_{35}+m_{45})/\sqrt{2} & (m_{45}-m_{35})/\sqrt{2} & 0\\
\end{smallmatrix}\right).
\]
\end{widetext}
We then refer to the  perturbation theorem for the eigenvalues of a matrix\cite{MatrixAnalysis} to get constraints on the eigenvalues of ${\mathcal{M}}^{\prime}$. 
Letting ${\mathcal{D}}=diag(a_{11},\cdots ,a_{nn}) \in Mat(n,\mathbf{R})$ and
 ${\mathcal{E}}=[e_{ij}]\in Mat(n,\mathbf{R})$, consider the perturbed matrix $ {\mathcal{D}}+ {\mathcal{E}}$. Then the eigenvalues of $ {\mathcal{D}}+ {\mathcal{E}}$ are 
 contained in the discs,
$$
\{z\in C : |z-\lambda -e_{ii}|\leq R^{'}_{i}({\mathcal{E}})=\sum_{\underset{j=1}{j\neq i}}^{n}|e_{ij}|\}\quad i=1,\cdots ,n,
$$
which are contained in the discs, 
$$
\{ z\in C: |z-\lambda_{i}|\leq R^{'}_{i}({\mathcal{E}})=\sum_{j=1}^{n}|e_{ij}| \}.
$$
Thus if $\hat{\lambda}$ is an eigenvalue of $ {\mathcal{D}}+ {\mathcal{E}}$, there is some eigenvalue $\lambda_{i}$ such that, 
$$
|\hat{\lambda}-\lambda_{i}|\leq |||{\mathcal{E}}|||_{\infty}.
$$
Here the matrix norm 
\begin{eqnarray}
|||{\mathcal{E}}|||_{\infty} &=& max  \lbrace m_{12}+ \sqrt{2} M_{14}, m_{12}+ \sqrt{2} m_{24}, \nonumber \\
& & \frac{M_{14}}{\sqrt{2}}+\frac{m_{24}}{\sqrt{2}}+\frac{(m_{45}+m_{35})}{\sqrt{2}}, \nonumber \\
& & \frac{M_{14}}{\sqrt{2}}+\frac{m_{24}}{\sqrt{2}}+\frac{(m_{45}-m_{35})}{\sqrt{2}},  \sqrt{2} m_{45} \rbrace . \nonumber
\end{eqnarray}
Thus the mass eigenvalues would be restricted to lie in  certain intervals determined in terms of the strengths of mixing parameters. 
In the absence of numerical data, we are not able to carry the analysis further\footnote{It is interesting to note that these bounds are independent of $M_{13}$. Then 
they would remain the same if the broken symmetry chain had an intermediate $SU(5)$ instead.}. 

\bigskip


\section{Summary and Concluding Remarks} \label{sec:conclusion}

We consider here an $E_6$ grand unified model with intermediate Pati-Salam symmetry
where the gauge bosons live in the adjoint representation $\{78\}$ of $E_6$ while  each generation of fermions 
are assigned to copies of the fundamental representation $\{27\}$. We take the Higgs bosons in $\{\bar{27}\}$ and 
in accord with the assumed symmetry breaking chain work out all physical, conserved $U(1)$ charges of the fermions. Furthermore 
the generic Yukawa couplings allowed by our symmetry breaking chain  are also pointed out.    
We concentrate our attention in particular on the electrically neutral fermion sector, neglecting here any horizontal mixing between different generations.
First of all we have active neutrinos of both chiralities. In order to differentiate between the masses of the left-handed and right-handed active neutrinos, a minimal seesaw mechanism is implemented by hand. Although such a mechanism wouldn't be allowed by our underlying GUT symmetry,
phenomenologically heavy right-handed active neutrinos are not observed, unlike the physical light left-handed active neutrinos. We further have 
a heavy exotic (Dirac) neutrino and a light (or even massless to the lowest order of approximation) sterile (Majorana) neutrino. Then we write down a
$5 \times 5$ real symmetric matrix that defines the masses and the mixing of these neutral leptons. We first discuss tree level masses by turning off all the mixing terms.
We note that when the mixings are turned on, the active and sterile neutrinos do not mix, while both of these types can mix with the exotic neutrinos on their own. We have shown  
up to second order perturbation approximation that the sterile neutrino mass would be inversely proportional to the exotic neutrino mass, and directly proportional to the 
product of the sterile-exotic mixing coefficients. 
The effects of the mixing of active and exotic neutrinos on the sterile neutrino mass are expected to show up at higher order perturbations.  
By the plausible assumption that exotic neutrinos would be the heaviest among all neutrino types, we also consider 
an expansion of the mass matrix around its maximum eigenvalue, taking the maximum to be the exotic neutrino mass $M_{34}$. 
Then we apply the perturbation theorem for matrices that allows us to put constraints on actual mass eigenvalues
in terms of the mixing parameters. The neutral lepton masses and mixings in a $E_6$ GUT with intermediate $SU(5)$ symmetry has been analysed before by 
Rosner\cite{rosner2014three}.
The $5 \times 5$ matrix of masses and mixing we derive here differs from Rosner's expression by the non-vanishing mixing parameter $M_{13} \neq 0$. This particular term enters into our 
mass eigenvalue expressions. Yet we haven't been able to pinpoint a physical process that would distinguish  between 
the two different symmetry breaking patterns. We note that the same single non-vanishing mixing parameter 
that differentiates between the two symmetry breaking patterns
also appears in the mass and mixing matrices of the colour triplet quarks and the charged leptons. 

The bosonic sector of SM consists of  8 (massless) gluons, 3 heavy weak intermediate bosons and one massless photon.  
Together with a single real scalar Higgs boson, this adds up to 28 physical bosonic degrees of freedom. 
The fermionic sector of SM  includes  3 generations each of  charged leptons, chiral neutrinos and iso-doublets of 3-coloured quarks. 
This adds up to  15 physical, fermionic degrees of freedom for each generation. 
In the $SO(10)$ GUTs,  fermions are assigned to the fundamental representation $\{16\}$. Therefore  to postulate the existence of a single heavy right handed neutrino 
over the SM fermion spectrum will be sufficient to fill up the whole multiplet\cite{pati2017}. 
On the other hand in the $E_6$ GUTs  fermions are assigned to $\{27\}$, so that one needs to postulate further the 
existence of heavy exotic fermions that amounts to 11 extra fermionic degrees of freedom for each generation. In the bosonic sector, on the other hand, the adjoint representation 
 $\{45 \}$  of $SO(10)$ asks for 18 extra intermediate vector bosons to be postulated while the adjoint representation $\{78\}$ of $E_6$ requires  51 new gauge bosons.  Thus, both of these models require one to postulate  the existence of large numbers of gauge bosons that would mediate lepto-quark interactions.  
 The fact that the fermionic sector of $SO(10)$ GUTs
are economical in that respect over the $E_6$ GUTs does not provide sufficient reason on its own to justify not to go beyond $SO(10)$ in a grand unification theory.   
Strong support for $E_6$ GUTs come from string theory motivated grand unification models\cite{harvey1981},\cite{nandi-sarkar},\cite{Lust-theisen1989},\cite{tye1999},\cite{itoetal2011}. 
For instance, effective heterotic string field theory models in 10-dimensions should contain either one of $SO(32)$ or $E_8 \times E_8$ gauge symmetry
groups for anomaly cancellation.
It is suggestive to consider the well-known chain of Coxeter-Dynkin diagrams 
for the E-series in the Cartan classification of complex semi-simple Lie algebras:
\begin{eqnarray}
E_8 \rightarrow E_7 \rightarrow E_6 \rightarrow E_5 &\cong &SO(10) \rightarrow  E_4 \nonumber \\
 &\cong & SU(5) \rightarrow SU(4) \rightarrow SU(3). \nonumber
 \end{eqnarray}
If we identify the $E_8$ at the upper end of this tower with one of the $E_8$'s in heterotic string theory and identify the $SU(3)$ at the lower end of the tower with the 
unbroken colour symmetry of  elementary particles,then it is natural not to stop at an $SO(10)$ GUT and  move over to $E_6$ GUTs towards $E_8$\cite{davis2015}.   

The electroweak unification  occurs at energy scales of $\sim 10^2 GeV$ at which the local (Abelian) gauge symmetry of QED is diagonally  embedded into a larger (non-Abelian) gauge symmetry group:
$$
U(1)_{QED} \hookrightarrow U(1)_{Y} \times SU(2)_{I}.
$$
The strong interactions on the other hand are described  in terms of quarks and gluons in QCD based on a symmetry group
$$
SU(3)_C \times G_F. 
$$
The flavour group $G_F$ accommodates a global horizontal symmetry among the quark generations. The local  colour symmetry group $SU(3)_C$ 
generates asymptotically free forces which might explain the observed confinement of colour. This is  a hypothesis yet to be theoretically verified. 
While going over to GUTs, moving up in the hierarchy steps at ever increasing energy scales, the unbroken colour symmetry $SU(3)_C$ is always  preserved. 
Therefore two big issues in grand unification schemes remain: (i) to find room for three and only three generations of fermions and (ii) to verify colour confinement hypothesis that is central to the success of QCD. $SO(10)$ GUTs are more economical in numbers but on both of the essential issues above $E_6$ GUTs seem more suggestive. 
An algebraic approach to colour confinement problem had been proposed long ago by G\"{u}rsey\cite{gursey1978}.
The fact that the structure of all exceptional Lie algebras including $E_6$  depends on the non-associtative properties of octonions is well-known\cite{Dray-Manogue}.
The 26 dimensional exceptional Jordan algebra of Hermitian $3 \times 3$ matrices over complex octonions plays a
unique role in the algebraic approaches to grand unification.
There are other related early work along these lines\cite{biedenharn1981},\cite{kerner}, yet this is still an open research direction.

\vskip 1cm
\bigskip

\acknowledgements
\ S.B. thanks  Ko\c{c} University for a Graduate Student Scholarship. 
We are grateful to Dr.Emre Mengi for discussions on matrix analysis.

\end{document}